\documentclass{article}
\usepackage{graphicx}

\input{tcilatex}

\begin{document}

\title{The Discontinuous Spatiality of Quantum Mechanical Objects }
\author{Helio V. Fagundes \\
%EndAName
Instituto de F\'{\i}sica Te\'{o}rica, Universidade Estadual Paulista\\
01405-900 S\~{a}o Paulo, SP, Brazil\\
E-mail: helio@ift.unesp.br}
\maketitle

\begin{abstract}
The quantum mechanical wave-particle dualism is analyzed and criticized, in
the framework of Reichenbach's concepts of phenomenon and interphenomenon.
It is suggested that the dual pictures be de-emphasized in the study of
quantum theory. In this connection a view of the electron and other such
particles as having discontinuous spatiality in time is presented and
discussed.
\end{abstract}

\section{INTRODUCTION}

Textbooks on quantum mechanics typically begin by discussing the
wave-particle dualism in its extreme aspects: such processes are best
described in the particle picture, such other in the wave picture. Some
texts, like Schiff's \cite{schiff}, for example, do then present a strictly
correct, but too succinct \textquotedblleft quantum mechanical
viewpoint.\textquotedblright\ After this the mathematical formalism is
developed, and only occasionally the qualitative aspects of the theory are
touched upon. It is left to the student the not-so-easy task of integrating
the qualitative and quantitative fundamentals of quantum theory into a
logically coherent, visualizable picture of the atomic world.

In the present paper I first make an analysis and criticism of the concept
of dualism. Our purpose is to sharply expose the artificiality and
non-uniqueness of both the wave and the particle pictures. To our knowledge
the best analysis of the quantum mechanical dualism is that of H.
Reichenbach, in his book \textit{Philosophic Foundations of Quantum Mechanics%
} \cite{reichenbach}. I am therefore guided by Reichenbach's ideas in our
study of the meaning of the wave-particle dualism.

Second, and more important, a qualitative picture of microscopic reality is
advocated, which is consistent with the more abstract interpretation of
quantum mechanics. As is well known, this interpretation assigns physical
reality to observable facts only. The key to our preferred language is to
abandon thinking of micro-objects -- I shall deal with \textit{electrons}
for definiteness -- as having permanent spatiality; that is, as having
spatial properties at all times. For instance, the question
\textquotedblleft Where is the electron when it is not
observable?\textquotedblright\ is meaningless, since the electron by
definition does not have a place when it does not participate in an
observable phenomenon. This meaninglessness is not a mere consequence of
unobservability. It is rather a matter of adequate definition.

In Section 2 Reichenbach's ideas of phenomenon and interphenomenon are
introduced, as a basis for the discussion of the wave-particle dualism. In
Sec. 3 an alternative language is presented, which pre-empties micro-objects
such as electrons or photons of classical-like spatial attributes, such as
temporal continuity of position or shape, trajectory, or propagation. I try
to make it clear that there is no real need for such attributes in
micro-objects. Next (Sec. 4) both viewpoints - i.e., those of Sections 2 and
3 - are applied to a physical example. I call the reader's attention to the
fact that physics progresses by increasing degrees of abstraction.
Concluding remarks are made in the last Section.

\bigskip

\section{REICHENBACH'S CONCEPTS OF \ PHENOMENON AND INTERPHENOMENON}

Reichenbach (p. 21 of \cite{reichenbach}) defines \textit{phenomena} as
\textquotedblleft occurrences which are so easily inferable from macroscopic
data that they may be considered observable in a wider
sense.\textquotedblright \footnote{%
I should perhaps add that a phenomenon leaves some trace of its occurrence,
which remains for some time, so that it can in every case be detected (at
least conceivably; see \cite{ayer}).}\ In the same page he rightly points
out that the phenomena are ... determinate in the same sense as the
unobserved objects of classical physics. As a matter of fact, physicists are
not troubled by the question of the reality of observable-but-unobserved
events, no more than by the question of the reality of unobserved trees. As
R. P. Feynman puts it: \textquotedblleft Nature does not know what you are
looking at, and she behaves the way she is going to behave whether you
bother to take down the data or not.\textquotedblright \cite{feynman} In
other words, the concept of phenomenon supposes \textquotedblleft Berkeley's
problem\textquotedblright\ solved. A satisfactory solution is given by A. J.
Ayer \cite{ayer}.

\textit{Interphenomena} are \textquotedblleft occurrences ... constructed in
the form of an interpolation within the world of
phenomena\textquotedblright\ (p. 21 of \cite{reichenbach}). They are logical
constructs similar to phenomena, usable to fill the gaps in the world of
phenomena, so that the whole have an appearance of continuity in time.
Notice that this is a purely mental process! It is precisely this process
that I intend to dispose of, as will be seen below.

Next, Reichenbach says that \textquotedblleft the distinction betIen
phenomena and interphenomena is the quantum mechanical analogue of the
distinction between observed and unobserved things\textquotedblright\ (p. 21
of \cite{reichenbach}). This analogy is somewhat misleading, because the
relation between phenomena and interphenomena is not so simple as that
between observed and unobserved macro-objects. If one likes labels, we might
call the question of interphenomena \textquotedblleft Berkeley's problem of
the second degree\textquotedblright : formerly one had to construct objects
out of actual or expected sense-experiences; here one has to define entities
related to experience only through its end points, the initial and final
events.

After these preliminaries, it may be said that the particle and wave
pictures in quantum theory represent different choices of interphenomena. In
the former, the gap between, say, two consecutive electron appearances (two
\textquotedblleft happenings,\textquotedblright\ or phenomena, involving the
electron) is filled by an imagined trajectory (interphenomena!) traveled by
the electron, imagined as a point-like object. With wave-like
interphenomena, on the other hand, the same gap if filled by imagining the
first phenomenon as a disturbance in a medium, then a wave-like propagation
of the disturbance, and finally a quick coalescing of the wave into another
point-like disturbance.

\bigskip

\section{THE IDEA OF DISCONTINUOUS SPATIALITY}

The considerations of the preceding Section will be illustrated by an
example in the next Section. Here the dualistic view is countered with what
I consider a more rational and satisfactory picture.\footnote{%
After completion of this work, my attention has been called to a possible
relation between the idea presented here and Margenau's concept of
\textquotedblleft latent observable." See Sec. 8.2 in \cite{margenau}.}

The point is this: there is no need for interphenomena. Beween two
consecutive, observable events nothing happens to the electron. During this
time interval the electron is nowhere, it is out of space. One is not
allowed to say that it has as \textquotedblleft indefinite
position\textquotedblright\ somewhere in the neighborhood of the events.
Contrarily to the situation with macroscopic things, micro-objects like
electrons do not constantly occupy a place, they have a place only when they
appear. Electrons have a permanent potentiality to interact, mathematically
expressed as its charge, and hence to appear in space, but no permanent
spatiality. (This is a qualitative content of the well known assertion that
the state function $\psi $ gives a probability amplitude, not a wave
amplitude. See also Section IV.3 in \cite{heisenberg}.) In other words, its
spatiality is limited, or discontinuous in time.I hope to make this point
clearer in next Section.

One might object that this picture has the disadvantage of being more
abstract than the dual pictures. My answer to this is: (1) \textquotedblleft
abstract\textquotedblright\ is a relative term: a moment's reflection may
convince the reader that the classical concept of trajectory is also quite
abstract, though less so than quantum mechanical concepts; and (2) progress
in physics is often reached by increasing degrees of abstraction, as for
example the abandonment of a material ether as the medium of electromagnetic
waves. This is only natural, since \textquotedblleft
concrete\textquotedblright\ ideas are just those of our too limited,
ordinary sense experience.

\section{DIFFRACTION OF AN ELECTRON}

Let us now illustrate the problem through an idealized\footnote{%
In practice electron diffraction is obtained by the passage of an electron
beam through a crystal.} experiment of diffraction of an electron by a slit
in a diaphragm \cite{reichenbach}.\bigskip \bigskip

\begin{figure}[h]
\centering\includegraphics[width=10cm,height=6cm]{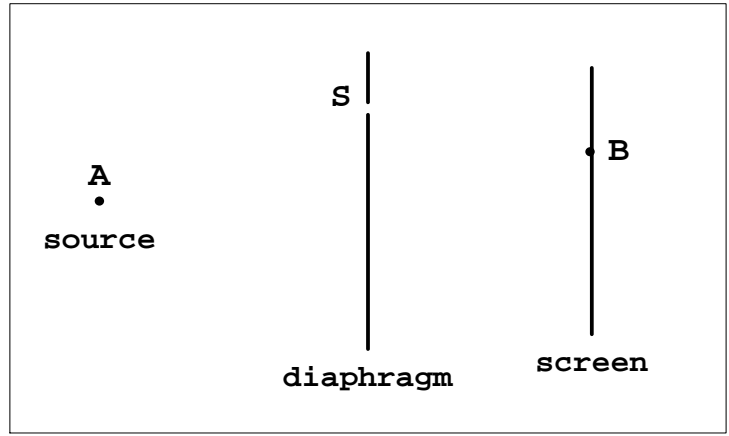}
\caption{Schematic view of electron diffraction}
\end{figure}

Figure 1 is a sketch of the experimental situation: \textit{A} is the
electron source, \textit{S} is the slit in the diaphragm, and \textit{B} is
the point on the screen where the electron is absorbed. The source is
supposed to be so weak that only one electron is emitted at a time. The
emission of an electron at \textit{A} and its absorption at \textit{B} are
point-like events, and are phenomena in the sense defined above. Let us call
them process $\alpha $ and process $\beta $, respectively. The space and
time gap between these phenomena has been filled by Reichenbach with
particle-like interphenomena and with wave-like interphenomena, which I
shall now discuss. Note that these constructs are not incorrect. They are
logically sound, indeed. The point to be emphasized in this paper is rather
their artificiality. My claim is that it is more illuminating\footnote{%
Despite Reichenbach's naming of the abstract views as \textquotedblleft
restrictive interpretations" (p. 33 of \cite{reichenbach}).} to discard
interphenomena altogether, and to consider quantum mechanical objects as
entities which do not always occupy space.

\subsection{Particle-like interphenomena}

In this picture the electron is said to be a concentrated lump of matter,
which after process $\alpha $ at \textit{A} travels through space along a
straight path from \textit{A} to \textit{S}. Then it proceeds along another
straight line from \textit{S} to \textit{B}, where process $\beta $ occurs.

This is called a \textit{normal interpretation}, which is another important
concept in Reichenbach's analysis: it means that the constructed
interphenomena obey physical laws which do not sharply differ from the laws
of the observed world of phenomena \cite{reichenbach}. (Otherwise it is
called an \textit{anomalous interpretation}. See below.)

Note that here the direction the electron will take after passing \textit{S}
can only be predicted in probability terms: this does not make the
interphenomena anomalous, since analogous situations exist in the world of
phenomena - for example, quantum phenomena too can only be predicted in
probability terms.

\subsection{Wave-like interphenomena}

This picture is less trivial, and more relevant to the present discussion.
We say that the electron is a wave-like disturbance of a medium, initiated
by phenomenon $\alpha $ at \textit{A} This is not an anomaly: the medium has
been invented for the support of interphenomena, but its ability to
oscillate is normal. Later the wave collapses in order to wholly pass
through slit \textit{S}, which is an anomaly: no such instantaneous collapse
exists for real waves. Then the wave propagates from \textit{S} to the
screen, and finally collapses again at \textit{B} for the production of
phenomenon $\beta $. So the wave-like interpretation is anomalous. But this
does not make it wrong. In Reichenbach's theory, it is as true as the
particle-like picture. Such anomalies are the price one pays for demanding
\textquotedblleft filling the gaps\textquotedblright\ between real events.

\subsection{The abstract picture}

This picture says that between phenomena $\alpha $ and $\beta $ nothing
happens to the electron. Needless to say, it has been assumed all along that
there is vacuum between the elements sketched in the figure. The $\psi $%
-function and the rules of quantum theory let us calculate the probability
of event $\beta $ after $\alpha $ has happened, but no spatial picture of
inter-events is allowed. The electron has position \textit{A} when $\alpha $
occurs and position \textit{B} when $\beta $ occurs. In between it does not
have either place or shape, it has no spatiality.

\subsection{Exercise}

It is left as an exercise for the reader to make similar discussion for what
is known in quantum mechanics as \textit{tunneling} through a potential
barrier \cite{schiff}. Especially noteworthy is the fact that its
particle-like interphenomena have the anomaly of energy non-conservation 
\cite{reichenbach}.

\section{CONCLUSION}

Reichenbach's analysis of the wave-particle dualism of quantum mechanics has
the great merit of showing that these interpretations are logical
constructs, neither unique nor unambiguous for a given physical situation.
While in a lower level - i. e., in the case of unobserved facts or objects -
similar constructs are very convenient and adequate, because unambiguous and
\textquotedblleft normal,\textquotedblright\ they are rather restrictive of
the imagination in the case of inter-events. I think it is more illuminating
to discard the concept of dualism, and to recognize that there is no need to
project all ideas of physical theory into space. Space, or more precisely
position in space, is in a sense produced by physical phenomena. The
macrophysical, persistent continuity of spatial forms is an illusion
analogous to many others, for example, the illusion of spatial continuity of
macroscopic matter. The proper language of physics is mathematical, and in
many cases mathematical statements do not allow for mapping on intuitive
space. Thus an electron is a linguistic entity logically similar to common
sense objects [4], but having a much poorer spatial character.\footnote{%
Or no spatiality at all; in this alternative view only events or phenomena
would have place. It is expedient, however, to make the electron and other
such `particles' appear in the spots where phenomena take place, which are
described as resulting from their interactions.}

Of course, the language of the wave and particle pictures has heuristic
value in problem solving and experiment setting. It is also a practical
jargon. We may even say that these pictures have some reality. However, the
actual measurements and calculations are expressed in mathematical language,
and physicists move so safely in this domain that they may not see how vague
are qualitative statements made in terms of the dual pictures.

I cannot propose that the idea of dualism be banned from physics. But I do
suggest that students pay more attention to the unrealism of considering an
electron as \textquotedblleft sometimes a particle and sometimes a
wave.\textquotedblright\ The electron is a mathematically defined thing,
which has the physical attributes of form and position only intermittently:
namely, when it participates in the production of a phenomenon.\newline

\bigskip

{\Large \textbf{ACKNOWLEDGEMENTS}}\newline
\newline
The Author wants to express his thanks to Professor P. Leal Ferreira, of
this Institute, for his critical reading of the manuscript, and to Professor
C. E. Bures, of the California Institute of Technology, for his
encouragement. This work was partly supported by FINEP.

\bigskip

\

\end{document}